\documentclass[conference]{IEEEtran}
\IEEEoverridecommandlockouts
\usepackage{cite}
\usepackage{amsmath,amssymb,amsfonts}
\usepackage{algorithmic}
\usepackage{graphicx}
\usepackage{textcomp}
\usepackage{xcolor}
\usepackage{multirow}
\usepackage[symbol]{footmisc}
\usepackage[colorinlistoftodos]{todonotes}

\def\BibTeX{{\rm B\kern-.05em{\sc i\kern-.025em b}\kern-.08em
    T\kern-.1667em\lower.7ex\hbox{E}\kern-.125emX}}

\newcommand\copyrighttext{%
  \footnotesize \textcopyright 2023 IEEE. Personal use of this material is permitted. Permission from IEEE must be obtained for all other uses, in any current or future media, including reprinting/republishing this material for advertising or promotional purposes, creating new collective works, for resale or redistribution to servers or lists, or reuse of any copyrighted component of this work in other works. 
  \linebreak Preprint submitted to the WSACC 2023 workshop, organized in the scope of the 9th IEEE International Smart Cities Conference 2023.
  }
\newcommand\copyrightnotice{%
\begin{tikzpicture}[remember picture,overlay]
\node[anchor=south,yshift=10pt] at (current page.south) {\fbox{\parbox{\dimexpr\textwidth-\fboxsep-\fboxrule\relax}{\copyrighttext}}};
\end{tikzpicture}%
}

\begin{document}

\title{A Study on Indoor Noise Levels in a Set of School Buildings in Greece utilizing an IoT infrastructure}
\makeatletter
% command to center 4th author in the title
\newcommand{\linebreakand}{%
  \end{@IEEEauthorhalign}
  \hfill\mbox{}\par
  \mbox{}\hfill\begin{@IEEEauthorhalign}
}
\makeatother

\author{
\IEEEauthorblockN{Georgios Mylonas\thanks{* Corresponding author: Georgios Mylonas, mylonasg@athenarc.gr}}
\IEEEauthorblockA{\textit{Industrial Systems Institute} \\
\textit{Athena Research \& Innovation Center}\\
\textit{\& Comp. Tech. Inst. \& Press ``Diophantus''}\\
Patras, Greece\\
ORCID: 0000-0003-2128-720X\\
}
%mylonasg@athenarc.gr}
\and
\IEEEauthorblockN{Lidia Pocero Fraile}
\IEEEauthorblockA{\textit{Industrial Systems Institute} \\
\textit{Athena Research \& Innovation Center,}\\
Patras, Greece\\
pocero@isi.gr}
\and
\IEEEauthorblockN{Stelios Tsampas}
\IEEEauthorblockA{\textit{Industrial Systems Institute} \\
\textit{Athena Research \& Innovation Center}\\
Patras, Greece\\
%ORCID: \\
tsampas@isi.gr}
\linebreakand
\IEEEauthorblockN{Athanasios Kalogeras}
\IEEEauthorblockA{\textit{Industrial Systems Institute} \\
\textit{Athena Research \& Innovation Center,}\\
Patras, Greece\\
ORCID: 0000-0001-5914-7523}
}

\maketitle

\copyrightnotice

\begin{abstract}
Monitoring noise pollution in urban areas in a more systematic manner has been gaining traction as a theme among the research community, especially with the rise of smart cities and the IoT. However, although it affects our everyday life in a profound way, monitoring indoor noise levels inside workplaces and public buildings has so far grabbed less of our attention. In this work, we report on noise levels data produced by an IoT infrastructure installed inside 5 school buildings in Greece. Our results indicate that such data can help to produce a more accurate picture of the conditions that students and educators experience every day, and also provide useful insights in terms of health risks and aural comfort. 
\end{abstract}


\section{Introduction}

In the past few years, the adverse potential effects of air and noise pollution on our society are being reconsidered. Recent estimates~\cite{maibach-2008} on the effect of noise pollution place it between 0.3-0.4\% to the GDP of the European Union. Moreover, recent reports by the European Environmental Agency (EEA) on noise levels~\cite{eea-report-noise-2020} strongly support this view. In this setting, countries have started taking noise pollution more seriously in terms of passing laws to deal specifically with this situation. As a characteristic example, the EU Environmental Noise Directive 2002/49/EC~\cite{noise-directive} (amended via Commission Directive (EU) 2015/996), instructs cities in the EU with certain characteristics to create noise levels maps. Furthermore, there are additional aspects to consider with respect to the aforementioned domains, e.g., taking into account the peculiarities of each site. As an example, although in Greece and Spain EU law and directives may be the same, the sources of noise pollution can be quite different in nature.

Apart from assessing the overall effect of noise pollution on our health and lives in general, cities have started investigating specific sectors of activity inside urban areas, in order to monitor and better understand the current situation, as well as create strategies and mechanisms to deal with this issue. In light of recent technological developments in smart cities and the IoT, it has also become much easier to implement noise monitoring systems. In this setting, in recent years we have seen many implementations using e.g., smartphones, to monitor noise levels inside city centers and other busy areas, utilizing the concepts of crowdsensing and citizen science. Overall, noise monitoring is a good example of why more fine-grained monitoring (e.g., through crowdsensing and citizen science) is needed in cities, in order to have a clear picture of how such phenomena evolve through time and under different settings. It can be a powerful means for helping us to better quantify the specific effect of noise pollution on our lives.

However, only recently we have begun to see the research community tackle the issue of noise pollution and comfort on areas outside of city centers, airports and ports. This has been facilitated by both inexpensive and relatively reliable IoT devices, as well as the rise of interest as regards health issues stemming from noise pollution. As an example, we have witnessed an interest regarding noise levels inside public buildings. Moreover, in terms of awareness, the scenarios for monitoring noise pollution have the potential to lead to an even deeper understanding of our environmental impact and to mobilize citizens to act immediately towards a more sustainable future. At the same time, we have also started to investigate its effect on very tangible aspects of our daily routine. In this spirit, in the context of the educational sector, recent findings~\cite{effects-noise-primary-school-children} indicate a correlation between noise in schools and learning performance.

In this work, we present a set of results regarding noise levels inside school buildings, an issue of great importance to the educational community and our society as a whole. We have utilized the IoT infrastructure available by the GAIA (Green Awareness In Action) Horizon 2020 project~\cite{gaia-website}, which installed IoT sensors in 25 school buildings in Europe. We have focused here on a set of schools in Greece, whose GAIA nodes were  equipped with noise monitoring sensors. Our results include interesting findings and also signal that it is feasible to develop such a solution with low cost and acceptable performance. Regarding the structure of this work, in Section 2 we briefly discuss previous related work, while in Section 3 we provide a short presentation of GAIA. In Section 4, we discuss the hardware and software setup utilized for noise levels monitoring in GAIA. In Section 5, we present a set of results produced by this IoT infrastructure and a discussion on these findings, and we conclude in Section 6.

\section{Related Work}

Noise levels monitoring is a popular field for crowdsensing, especially by using smartphones. Previous work focused on validating the crowdsensing approach, with a gradual shift in recent years to including additional calibration aspects, in order to have more reliable data. \cite{Guo2015MobileCS} provides a comprehensive of the mobile crowdsensing field, while \cite{pham-smartsantander} is an early example of using SmartSantander for urban noise monitoring. Since noise and sound in general has a tremendous effect on human psychology, this aspect of the project offers many paths for exploration. Especially in terms of the interplay between soundscapes and our everyday lives, there are numerous scenarios to explore. The potential of an acoustics/noise study of school buildings is even more interesting given the connection between student performance and overall comfort inside classrooms, as well as health concerns for educators. 

SONYC~\cite{SONYC} (Sounds Of New York City) is an example of a system enabling city-wide noise levels monitoring, using inexpensive IoT platforms and machine learning to categorize the various sources of noise inside urban areas. Other works like \cite{radicchi-1} and \cite{radicchi-2} focus on aspects of involving citizens more actively in noise pollution monitoring and sound mapping inside urban areas, utilizing co-creation concepts and techniques.

Regarding hardware designs, in the past few years we have seen several designs specifically for noise levels monitoring. This is especially important, since the domain of audio and acoustics traditionally has been using rather expensive hardware solutions and processes for such applications, e.g., creating official noise maps. Furthermore, there are currently available open source options like Audiomoth~\cite{audiomoth}, which can significantly simplify the overall process for developing a noise levels-related IoT application.

In recent years, we have seen several studies focusing on noise levels in educational environments. \cite{effects-noise-primary-school-children} studied the effect of noise on the performance of primary school students in England and Wales and found that external noise events have a significant impact on their performance. \cite{finland-school-noise} presented results regarding activity noise and acoustics in Finnish school buildings, and concluded that very few classrooms fulfilled the respective criteria set of the Finnish national standard. %\cite{acoustic-comfort-learning-spaces}
%\cite{survey-uk-secondary-schools} 
Our work differs from previous approaches in facilitating a more systematic and low-cost approach to monitoring noise levels inside schools, which could also enable long-term studies of the effects of noise in such environment on the performance of students. It could also facilitate the production of related datasets, which are thus far relatively scarce.

\section{The GAIA Project}

The work presented here was conducted in the context of the GAIA project \cite{gaia-ieee-pervasive}. GAIA aimed at increasing sustainability awareness and promoting responsible behavior towards energy  efficiency in educational communities. IoT technologies in the educational  buildings involved in GAIA are used to realize data-driven educational activities, instead of implementing invasive solutions and retrofitting schools with actuators. Sensors were installed in strategic parts of the schools to help students monitor the energy consumption of their building, or specific rooms, and become aware of the impact of environmental parameters (e.g., building/room orientation) and personal behaviour into the schools' energy consumption profiles. In the framework of GAIA, this type of activities were carried out, with proper adaptation, in schools located in Greece, Italy  and  Sweden of different education levels (i.e., primary to high school), while also covering different climatic conditions, cultural habits and regulatory frameworks.

In order to implement the IoT-driven flavour in GAIA's educational activities, sensors were deployed in the involved schools, thus giving rise to a real-world multi-site IoT testbed. This IoT infrastructure, hereafter referred to as the GAIA IoT Platform, comprises over 1200 IoT monitoring  endpoints  and it is made of heterogeneous hardware and software technologies, leveraging different hardware/sensor vendors, as well as open-source solutions \cite{hardwarex}. In each school, a set of sensors were deployed to monitor the following parameters: active power, energy, internal  environmental parameters (e.g., luminosity, noise), external weather and pollution parameters. The actual deployment at each site is custom and adapted to the features of the building (e.g., size, number of students, orientation) and, in some cases, to the availability of already installed sensors. Measurements are continuously acquired by the GAIA IoT Platform~\cite{gaia-sensors} which implements data processing services. Sensor data streams are processed and aggregates are extracted in real time at different time granularity (e.g., 5-minute, hour, day, etc.). These data can be accessed through REST APIs, and a set of applications have been developed to facilitate students and teachers in accessing and using sensors data. 

In the framework of GAIA, a set of IoT-driven educational activities have been designed and experimented that exploit sensors data, gamification and IoT hands-on experience. Gamification-inspired activities have been centered around the GAIA Challenge, a serious game web application, while hands-on IoT activities leveraged the GAIA IoT educational lab kit, made of hardware and software technologies supporting the design of IoT classes/tutorials.

\section{Methods}

\subsection{Hardware and Software utilized}

Regarding the hardware sensors utilized in the GAIA project, due to its multi-year span in deployment across schools, several sensors were used, with variable levels of capabilities, accuracy and cost. Thus far, the following 4 sensors have been used to monitor noise levels:

\begin{enumerate}
\item Openjumper OJ-CG306~\cite{openjumper}
\item Grove Sound Sensor~\cite{grove-sound}
\item Sparkfun Sound Detector~\cite{sparkfun-noise}
\item Gravity analog Sound Level Meter (SEN0232)~\cite{sen0232-sound-sensor}
\end{enumerate}

The first three sensors that were used for noise monitoring were not really orientated to measure sound, since they were oriented to detect noise thresholds rather than noise levels, but they provided some basic capabilities. Since their cost was very low, it also help to keep the overall cost of the IoT nodes down, and we tried to compensate for the lack of hardware capabilities with software solutions. Since all 3 sensors were not calibrated, we utilized a crude calibration approach described below. However, we do not claim any kind of theoretical accuracy when we use them as a decibel meter, because that is not their original purpose. The fourth sensor (SEN0232) is the only one that provides output as decibels, has a reasonable measuring range (30dBA-130dBA) and error (±1.5dB), as well as frequency response (31.5Hz--8.5KHz).

Regarding calibration for the Openjumper OJ-CG306 sensor, it has a set of sensitivity and gain parameters. The sensor detects the noise level in  two  stages: (1) we capture noise samples using this electret microphone characterized by a specific sensitivity, and, (2) we use an amplifier to provide output that can be read using an analog input. In general, the sensitivity of an analog microphone is typically specified in logarithmic units of dBV (decibels with respect to 1V), giving an estimate of how many Volts the output signal is for a given sound pressure level. For this specific sensor, 50dBV correspond to a sensitivity of $S=316.22V/Pa$, while the preamplifier was set to a gain of 26dB ($G=19.95$) using the onboard potentiometer available. Such low-cost sound pressure level sensors are generally used to detect the presence of sound in the environment, but in our software we calculate the sound level in dB by processing multiple samples of the analog output of the sound pressure level sensor. To calculate the absolute sound pressure level $L_p$ (or SPL), we first use the measured $V_{rms}$ to calculate the $P_{rms}$ in Pa using:

\begin{equation}
P_{rms} = (V_{rms}/G)/S
\end{equation}
          
We then calculate the $L_p$ in dB using the following equation, with $P_{ref}$ being the reference level of SPL (the threshold of SPL for the human voice). 

\begin{equation}
    L_p = 20log_{10}(P_{rms}/P_{ref}) = 20log_{10}P_{rms} - 20log_{10}P_{ref}
\end{equation}

For the Grove Sound Sensor, since it does not provide Gain and Sensitivity, we calculated their correlation with a calibrated decibel meter, and then calculated a linear regression between them. Note that we sometimes need regression lines different for every device.

For the Sparkfun Sound Detector Sensor, we utilized the same logic. The regression function  used was the following:

\begin{equation}
    (int) newStatus = (int)(20*log_{10}((value/20))+50) 
\end{equation}

\begin{equation}
    newStatus = 0.12 * maximum + 44.0     
\end{equation}

Again, we sometimes need custom regression lines for each device. However, for the Analog Sound Level Meter (SEN0232), there was no need to use such methods, since it produces directly noise level values in decibels that are close to the values produced by a reference trusted decibel meter for the same inputs, which we used to make comparisons.

%The sound pressure $p_{rms}$ is a measured root mean square (r.m.s.) value and the internationally agreed reference pressure $p_{ref}=2\times10^{-5}Pa$. When this value is substituted into the equation\ref{eq:9}, the convenient alternative form is obtained (equation \ref{eq:10})   
%\begin{equation}
% L_{p}=20log_{10}(p_{rms})+94(dB)
% \label{eq:10}
%\end{equation}
% 
%In order to compute the root-mean-square (RMS) amplitude $v_{rms}$ the instantaneous voltage output are squared, averaged and the square root of the average is taken \ref{eq:11}. $N=32$ samples are used to estimate each RMS value.      
%
%\begin{equation}
% v_{rms}= \frac{1}{N}\sum_{n=0}^{N-1}V_o[n]^2 (V)
% \label{eq:11}
%\end{equation}
%
% The upper processing is made with the direct analog readings (values from 0 to 1024) of the SPL module output signal and then convert to Voltage $v_{rms}$. Next, the sound pressure $p_{rms}$ is calculated using Sensitivity and Gain \ref{eq:12}  

%Replacing this value on the equation \ref{eq:10} a good approach to instantaneous SPL is estimated. On the cloud is upload the middle value of all all the  reading each upload period. 

In Fig.~\ref{fig:devices-collage-1}, examples of the actual devices utilized in GAIA schools can be seen. Due to the fact that the deployment of the devices took place over a period of 3 years, there were several changes to the hardware utilized in these boards, which are also reflected on the use of the 4 hardware modules mentioned above. Fig.~\ref{fig:devices-collage-2} provides some examples of the actual installation of such IoT nodes in classrooms of several schools. These nodes were installed above a certain height, in order to make it more practical and safer to use them in school environments, since several of the buildings belonged to primary schools. In most cases, the nodes were installed near the desks of the educators inside classrooms.

\begin{figure*}
  \includegraphics[width=\linewidth]{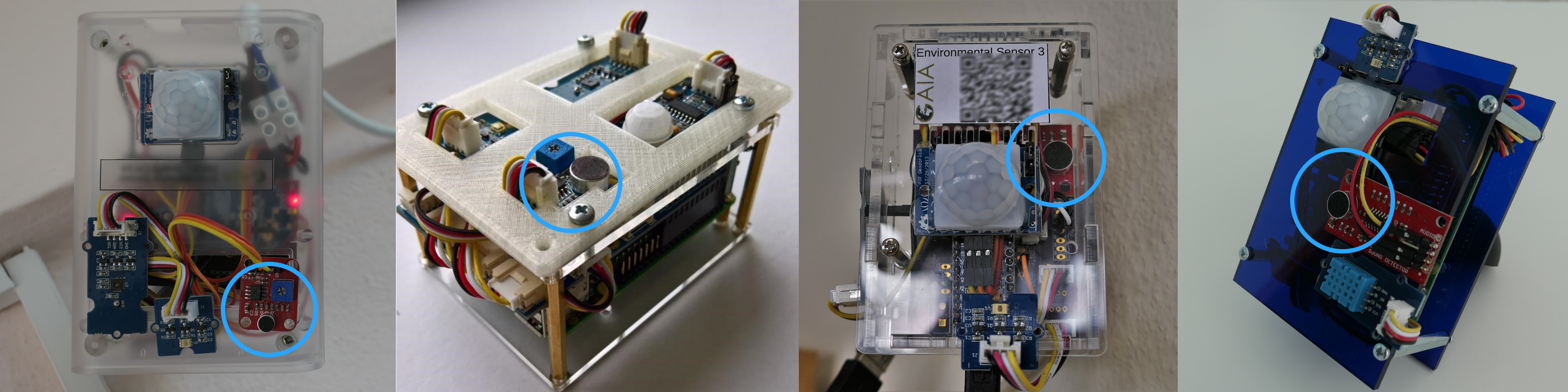}
  \caption{Some examples of the various revisions of the IoT nodes used in the school buildings of GAIA.The sound-related sensing components are circled to highlight their location on the nodes.}
  \label{fig:devices-collage-1}
\end{figure*}

\begin{figure*}
  \includegraphics[width=\linewidth]{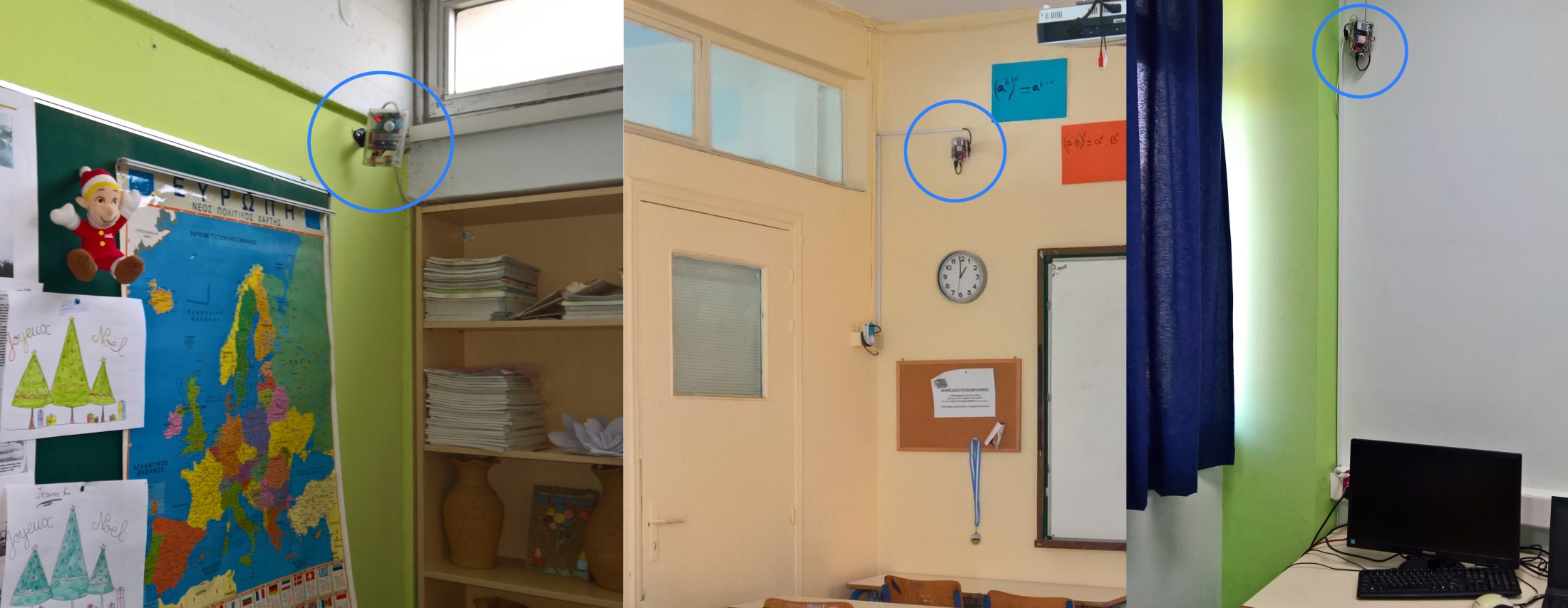}
  \caption{Some examples of installations of the IoT nodes inside the school buildings participating in the GAIA project, with the IoT nodes circled to pinpoint their location. The nodes were installed above a certain height to prevent damage or accidents, as well as minimize acting as a visual distraction to students.}
  \label{fig:devices-collage-2}
\end{figure*}

Regarding privacy aspects, all data produced by the sensors are processed directly on the IoT nodes in order to produce average values for noise levels every 30 seconds or more (e.g., 1-minute averages), depending on the available resources. Only then are measurements sent to the rest of the infrastructure of GAIA for further processing and storage, minimizing privacy risks, since we are dealing only with such average noise levels data and not actual sound recordings.

\subsection{Timeline of Data Collection}

Regarding the timeline of the data collection, as mentioned above, GAIA took place over the course of 3 years and more specifically between February 2016 and May 2019, while the infrastructure is still operational to a certain extent. In Table~\ref{tab:data} in the following section we have included data mostly from the school year 2018-2019. It should be noted that the COVID-19 pandemic has seriously affected data in the following school years. However, we also include an example of more recent data (May 2021), which were gathered over a short time period.

\section{Results - Discussion}

In this section, we present some results with respect to noise levels inside 4 schools in Greece, with 3 of them being primary schools and 1 a Junior High school. In these schools, we installed, along with more typical environmental sensors, nodes with digital noise level meters. These nodes have been calibrated to a certain degree before installation. They essentially calculate the average noise level every 5 minutes, and then report this numbers to GAIA’s cloud infrastructure. In the following table, we present our results for multiple rooms in each school that represent the percentage of working hours in these schools in which the noise levels exceed 40, 50, 60, 70, 80 and 85 dBA.

% \usepackage{multirow}
\begin{table*}[]
\centering
\begin{tabular}{|l|l|l|l|l|l|l|l|l|}
\hline
School & Room &  \% $>$ 40dBA & \% $>$ 50dBA & \% $>$ 60dBA &  \% $>$ 70dBA & \% $>$ 80dBA & \% $>$ 85dBA & Time period \\\hline
\multirow{4}{*}{Primary School A} & 1 & 35.55\% & 27.26\% & 22.49\% & 18.62\% & 3.65\% & 0.28\% & 08:00-16:00 \\\cline{2-9}
                  & 2 & 42.86\% & 37.23\% & 28.7\% & 8.43\% & 0.97\% & 0.02\% & 08:00-16:00 \\\cline{2-9}
                  & 3 & 48.32\% & 41.29\% & 32.09\% & 8.14\% & 0.16\% & 0.00\% & 08:00-16:00 \\\cline{2-9}
                  & 4 & 44.15\% & 33.13\% & 27.47\% & 7.41\% & 0.86\% & 0.16\% & 08:00-16:00 \\\hline
\multirow{4}{*}{Primary School B} & 1 & 67.82\% & 48.53\% & 38.31\% & 16.89\% & 5.77\% & 2.72\% & 08:00-16:00 \\\cline{2-9}
                  & 2 & 28.04\% & 0.0\% & 0.0\% & 0.0\% & 0.0\% & 0.0\% & 08:00-16:00 \\\cline{2-9}
                  & 3 & 13.21\% & 0.21\% & 0.0\% & 0.0\% & 0.0\% & 0.0\% & 08:00-16:00\\\cline{2-9}
                  & 4 & 20.2\% & 0.3\% & 0.02\% & 0.0\% & 0.0\% & 0.0\% & 08:00-16:00 \\\hline
\multirow{5}{*}{Junior High School C} & 1 & 55.04\% & 50.2\% & 42.01\% & 22.84\% & 2.19\% & 0.31\% & 08:00-14:00 \\\cline{2-9}
                  & 2 & 53.75\% & 46.91\% & 38.15\% & 20.62\% & 3.63\% & 0.89\% & 08:00-14:00 \\\cline{2-9}
                  & 3 & 46.49\% & 35.22\% & 22.26\% & 3.29\% & 0.0\% & 0.0\% & 08:00-14:00 \\\cline{2-9}
                  & 4 & 84.63\% & 46.89\% & 38.65\% & 27.01\% & 5.9\% & 1.95\% & 08:00-14:00 \\\cline{2-9}
                  & 5 & 10.22\% & 0.02\% & 0.0\% & 0.0\% & 0.0\% & 0.0\% & 08:00-14:00 \\\hline
\multirow{5}{*}{Primary School D} & 1 & 77.76\% & 62.47\% & 51.48\% & 28.72\% & 2.12\% & 0.31\% & 08:00-14:00 \\\cline{2-9}
                  & 2 & 76.9\% & 48.32\% & 39.89\% & 17.53\% & 0.47\% & 0.03\% & 08:00-14:00  \\\cline{2-9}
                  & 3 & 79.36\% & 49.56\% & 41.42\% & 24.81\% & 1.65\% & 0.23\% &  08:00-14:00\\\cline{2-9}
                  & 4 & 70.19\% & 50.97\%  & 41.23\% & 18.56\% & 1.05\% & 0.07\% & 08:00-14:00 \\\cline{2-9}
                  & 5 & 54.06\% & 46.98\% & 36.94\% & 9.08\% & 0.14\% &0.0\%  & 08:00-14:00\\\hline
\end{tabular}
\caption{Noise levels measurements from 4 schools in Greece. For each room monitored in each school, the percentage for exceeding a certain threshold (40, 50, 60, 70, 80, 85 dBA) is shown, along with the respective time periods for which the schools were open.}
\label{tab:data}
\end{table*}

\begin{figure*}
  \includegraphics[width=0.95\linewidth]{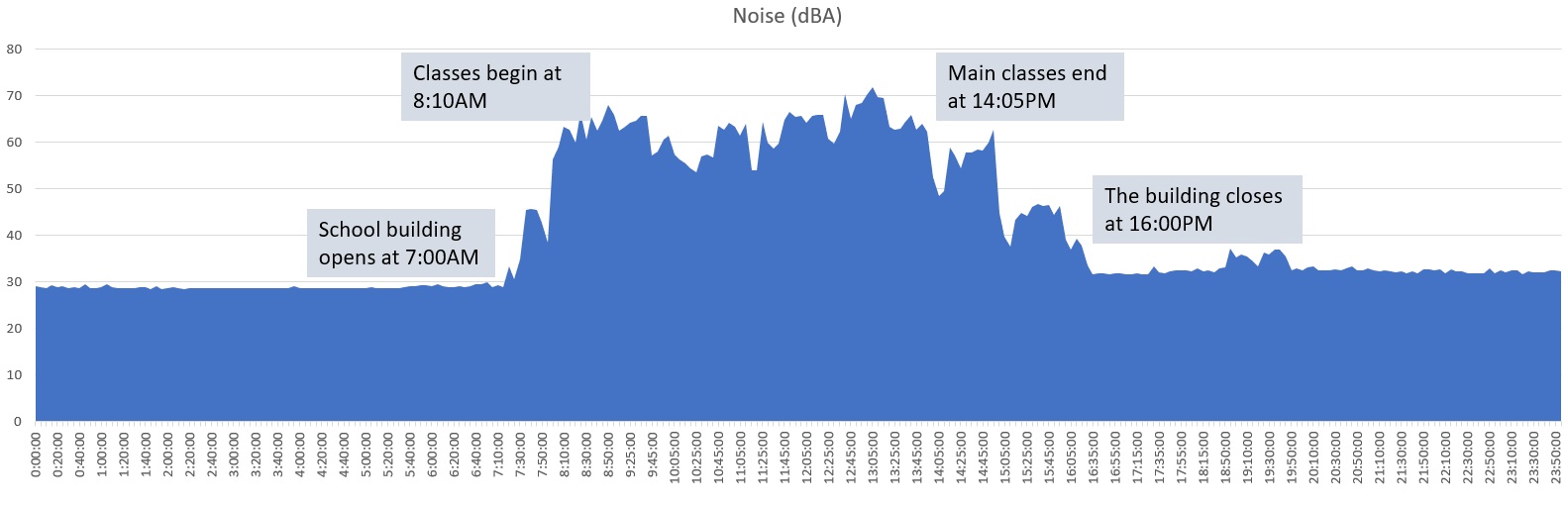}
  \caption{An example of the data retrieved for the average noise levels inside a school building over the course of 24 hours during a week day.}
  \label{fig:noise-example-day}
\end{figure*}

The World Health Organization and the European Union have made public a set of guidelines~\cite{who} with respect to noise. The most well-known of these guidelines is the one suggesting that people should not be continuously exposed to noise above 85dBA for over 8 hours. However, for children this value is closer to 70dBA, while there exist other guidelines for schools that state that noise inside schools should not exceed 40dBA.

From Table~\ref{tab:data}, we  see that in these schools even the threshold of 85dBA is reached, although not for prolonged periods. E.g., in Room 1 of School B, there is on average a noise level of above 85dBA for around 14.5 minutes per day, which, although not damaging, is certainly concerning. As we move to lower noise level thresholds, we see much longer time periods on average. There are rooms in these schools in which noise levels are above 70dBA for more than 1 hour on average every day. This is an issue that requires further investigation and actions to inform students and teachers, as well as the public about the dangers involved when dealing with such high noise levels.

Moving on to a more specific example of the data captured inside a school building, Fig.~\ref{fig:noise-example-day} captures the average noise levels inside a primary school, i.e., from all classrooms monitored, during a 24-hour period in May 2021. By using such data, it is quite straightforward to detect the exact time when school activities start and end, while it is also easy to spot periods when noise levels start to become an issue for the educators and the students inside classrooms. Noise levels are stable until 7:00AM, when the building opens and staff start to come in, with the students arriving at around 8:00AM. Classes begin at 8:10AM and continue until 14:05PM, after which a subset of the students and staff stays inside the building until 16:00PM.

In this case, we see that, at least in terms of average values, noise levels do not exceed 70dBA more than 5-10 minutes, and this happens near the end of the school day. We also see that there is an average noise level of 65dBA for much of the day. Some noise level readings between 18:30 and 19:50, which are a bit above the , could indicate activities by third groups utilizing the facilities of the school building when no educational activities are carried out, which is a common practice in Greece.

Furthermore, we can see that there exist quite significant differences between the indoor noise levels recorded, even for this limited set of school buildings. This could be seen as further proof that the situation with respect to indoor noise pollution inside public buildings like schools depends on a number of parameters, such as its location, construction, or factors like the social composition of the users of the building or the type of school (e.g., educational level and orientation).

Another benefit of our approach is that it can be used to assess the level of acoustic comfort during off-school hours, as well as in unoccupied classrooms during the day. This could be utilized to monitor the levels of outdoor noise entering the building, thus evaluating to a certain degree the noise insulation of the building when combined with noise level sensors installed outside the building to enable the comparison. Moreover, this approach could also enable further studies to associate student performance with noise levels, or study the effect of noise on physiological parameters like heart rate, stress, etc.

\subsection{Limitations}

Regarding limitations of this work, as mentioned, the hardware used was not calibrated to professional standards, thus leaving a lot of room for improvement as regards this aspect. %When comparing the latest hardware revision of our nodes with the results from older ones, whose primary purpose was not noise levels monitoring, we saw that the results from the old nodes in general correlate with the new ones reasonably well. 
Moreover, the positioning of the nodes inside the building/classrooms also affects monitoring, since we tend to measure noise not at e.g., the position of an educator in the classroom, but at a distance of some meters (1 to 3 meters in most cases) from that point. Thus, further work is required to ensure the correctness of the results produced. We have also monitored only indoor noise and do not take into account outdoor noise sources and their effect on indoor noise levels.
% add something here about missing values

\section{Conclusions}

In this work, we presented briefly our implementation of an IoT platform for noise monitoring inside schools, in the context of a research project. Our results indicate that by using such a system, it is feasible to capture the realities of indoor noise in this sensitive environment. Given the nature of such environments and the conditions in which educators and students operate each day, we believe such systems could give interesting insights to local governments and education ministries, in order to detect any potential issues. Regarding our future work, we intend to combine the indoor readings with ones from sensors located outdoors, as well as investigate the noise levels inside additional school buildings and correlations with academic performance data. We also intend to experiment with noise levels-based educational activities in the schools collaborating with us, as a means to engage with students and educators and inform them about issues linked to indoor noise levels.

\section*{Acknowledgment}

This work has been partially supported by the ``European Extreme Performing Big Data Stacks'' project, funded by the European Commission (EC) under the Horizon 2020 framework and contract number 780245, and the ``Green Awareness In Action'' project, funded by the European Commission and the Executive Agency for Small and Medium-sized Enterprises (EASME) under the Horizon 2020 framework and contract number 696029. This document reflects only the authors' views and the EC and EASME are not responsible for any use that may be made of the information it contains.

\bibliographystyle{IEEEtran}
\bibliography{bibliography}

\end{document}